\documentclass[12pt]{article}
\usepackage{graphicx}

\title{ Odderon and Pomeron from the Vacuum Correlator Method}
\author{  A.B.Kaidalov, Yu.A.Simonov\\
 State Research
Center\\Institute of Theoretical and Experimental Physics, \\
Moscow, 117218 Russia}
 \date{}
\newcommand{\beq}{\begin{eqnarray}}
 \newcommand{\eeq}{\end{eqnarray}}
\newcommand{\be}{\begin{equation}}
 \newcommand{\ee}{\end{equation}}

 \def\la{\mathrel{\mathpalette\fun <}}

\def\fun#1#2{\lower3.6pt\vbox{\baselineskip0pt\lineskip.9pt
\ialign{$\mathsurround=0pt#1\hfil ##\hfil$\crcr#2\crcr\sim\crcr}}}

\newcommand{{\SD}}{\rm SD}

\newcommand{\ver}{\mbox{\boldmath${\rm r}$}}

\newcommand{\vep}{\mbox{\boldmath${\rm p}$}}

\newcommand{\veR}{\mbox{\boldmath${\rm R}$}}

\newcommand{\vexi}{\mbox{\boldmath${\rm \xi}$}}
\newcommand{\veta}{\mbox{\boldmath${\rm \eta}$}}

\begin{document}
\maketitle

\begin{abstract}
Glueball masses with $J\leq 7$ are computed both for $C=+1$ and
$C=-1$ using the string Hamiltonian derived in the framework  of
the Vacuum Correlator Method. No fitting parameters are used, and
masses are expressed in terms of string tension $\sigma$ and
effective value of $\alpha_s$.

We extend the calculations done for $J\leq 3$ using the same
Hamiltonian, which provided glueball masses in good agreement with
existing lattice data, to higher mass states. It is shown that
$3^{--}, 5^{--}$ and $7^{--}$ states lie on the odderon
trajectories with the intercept around  or below 0.14. Another
odderon trajectory  with $3g$ glueballs of $Y$-shape,  corresponds
to 11\% higher masses and low intercept. These findings are in
agreement with recent experimental data, setting limits on the
odderon contribution to the exclusive $\gamma p$ reactions.

 \end{abstract}

\section{Introduction}
There was a renewal  of interest in glueballs recently, mostly
connected to pomeron and odderon trajectories \cite{o1}-\cite{o6}.
High-spin glueballs, e.g. $4^{++}, 6^{++}$ were calculated on the
lattice \cite{o1,o2}, and the problem of odderon attracted much
attention \cite{o3,o4,o5}. Search for odderon exchange at HERA
\cite{o6,o7}
 in the reactions $\gamma p \to \pi^0 p, \gamma p \to \pi^0 \pi^0 p$,...
   has not provided any indications for an existence
 of these processes at high energies\footnote{Note that the usual
 $\omega, \rho$ -- Regge poles with  $\alpha_u(0) \approx \alpha_\rho
 (0) \approx 0.5$ give  contributions to these reactions, but  the
 corresponding cross sections decrease with energy as
 $\frac{d\sigma}{dt}|_{t=0} \sim \frac{1}{5}$.} and has lead to
 rather strong limits on the  corresponding cross sections. This
 allows to put limits on the intercept of the odderon in the
 models which predict couplings of the odderon with hadrons
 \cite{o4}. Study of this problem
  allows to distinguish between the
models and seriously question some of them. Therefore it seems to
be necessary to make a more detailed analysis of our theoretical
calculations \cite{o8} and to compare additional high-spin
glueball masses.

A theoretical study of
 glueballs in QCD was started in \cite{7}-\cite{10}
 and is closely related to the
 problem of the   pomeron, i.e. leading Regge pole, which
 determines the asymptotic behavior of scattering amplitudes at very
 high energies.  It is usually assumed that the  pomeron in QCD is
 mostly gluonic object \cite{11} and glueball resonances with vacuum
 quantum numbers and  spins belong to this trajectory. Another
 interesting hypothetical Regge singularity is the "odderon", which
 has negative signature and $C$ parity and can be built out of at
 least 3 gluons.  Most studies of the pomeron and odderon
 singularities in QCD are based on applications of the perturbation
 theory \cite{12}.

Our method is based on the QCD path integral formalism, where all
dynamics is encoded in field correlators -- the so-called Field
Correlator Method (FCM) \cite{o15} (for a review see \cite{o16})
starting from that one can derive in the limit of small gluon
correlation length $\lambda$ the relativistic  Hamiltonian
\cite{o17}, which effectively describes the fundamental (adjoint)
string with quarks (gluons) at its ends. In this simple limit,
$\lambda\to 0$, Hamiltonian is local but nonlinear in $\vep^2,
\hat L^2$, which reflects complicated dynamics of relativistic
rotating string (similar results are obtained in \cite{o18}).

One should stress, that nonperturbative  (NP) approach started in
\cite{o19} and  developed in \cite{o8}, is based on the
extrapolation of trajectory $J(t)$  connecting  $2^{++}, 4^{++},
6^{++}$ states ( for pomeron) and $3^{--}, 5^{--}, 7^{--}$, states
(for odderon) to the physical region of scattering, $t=M^2\leq 0$.
In doing so one assumes that no extra singularities  appear is the
$J$ plane on the way to $t=0$.  In case of pomeron indeed two
other trajectories, $f$ and $f^{(\prime)}$ exist which intersect
with glueball trajectory and therefore one should takes into
account mixing between them. As  a result in \cite{o8} a combined
trajectory was calculated with realistic pomeron intercept. In
this approach perturbative contributions are of subsidiary
character and can shift the intercept by approximately 0.2.

It is important to stress that the region $t\approx 0$ as well as
$t\approx M^2>0$, $ M\sim$ several GeV belongs to the primarily
nonperturbative   regime, where all  intergluonic and interquark
distances are large, hence the extrapolation of trajectories to
$t=0$ is inside the NP domain.

In the case of odderon in \cite{o8}  the intersection of the
leading gluonic trajectory with $C=-1$ with $q\bar q$ trajectories
does not take place, since $\rho, \omega$ trajectories have much
larger intercepts, and the resulting  odderon  intercept in
\cite{o8} obtained using the $3^{--}$ glueball mass and assumed
slope of $(2\pi\sigma_{adj})^{-1}$ was predicted around -1.5. This
value excludes possible odderon discovery in reactions $\gamma
p\to (\pi^0, f_2^0, a_2^0)X$ \cite{o6,o7} within the upper limits
set by the authors.

On the other hand, the perturbative (BFKL) approach \cite{12}
starts from another premises. It considers pomeron (odderon) as
$2g$ (3g) system of Reggeized gluons  having only perturbative
gluon exchanges, which is justified in the perturbative  domain of
small $(\la 1$ GeV$^{-1}$) interparticle distances, the situation
which might be  realized for large negative $t$, large $s$, and
very small sizes of color dipoles exchanging pomeron. Therefore
one encounters  the problem  of analytic continuation of pomeron
singularity to the nonperturbative domain of $t=0$ (and realistic
dipole sizes).

Assuming this can be done, one obtains both pomeron and odderon in
vicinity of $J=1$ \cite{12},~~ see \cite{o20}, \cite{ o21} ~and
~\cite{o5}~ for ~more ~discussion~ of ~~odderon.

While for pomeron this is reasonable (however the coincidence of
pomeron  with $J=1$ is trivial  in the limit of small $\alpha_s$),
for odderon it presents a prediction of strong  negative $C$ --
parity--odd contribution to different reactions, in particular of
the type mentioned above \cite{o6,o7}.

In a different approach(see \cite{o4} and refs. therein) a model
nonperturbative picture for high-energy scattering was exploited
yielding  the odderon intercept $\alpha_{odd} (0)=1$, which as
well as the BFKL prediction is at odds with existing data.

Therefore we feel it is necessary to clarify the situation with
glueball trajectories and to this  end to extend our previous
calculations \cite{o8} to higher spin states, namely we calculate
the masses of glueballs with $C=+1, J=0^{++}, 2^{++},4^{++},
6^{++}$ and with $C=-1, J=1^{--}, 2^{--}, 3^{--}, 5^{--}, 7^{--}$.
From those we calculate the odderon and pomeron trajectories  and
find the appropriate intercepts.

Another aim of our calculations is the comparison of newly found
glueball masses with existing lattice data
\cite{o1,o2,o22,o23,o24}. In our previous work \cite{o8}  we have
found a good agreement of all glueball masses with lattice data.
Recently new states, $4^{++}$ and $6^{++}$ have been computed
\cite{o1,o2} and we can compare those with our analytic results.
We also compare our results with recent analytic calculations
\cite{o3, o25, o26}. The paper is organized as follows. In section
2 the Hamiltonian is given together with spin terms following
\cite{o8} and masses for $gg$ glueballs are calculated.

In section 3 two possible configurations of 3g glueballs  are
defined, a $\Delta$--type and an $Y$--type, and the resulting
masses are calculated. In section 4 glueball trajectories are
obtained and intercepts are found and discussed. The concluding
section 5 is devoted to the discussion of the results from the
point of view of lattice and experiment correspondence, and to
possible improvements.

\section{String Hamiltonian and spin corrections}

The Hamiltonian for the $gg$ system was derived in \cite{o8} in
the same way as it was done for the $q\bar q$ system \cite{o28},
and can be written as

\be H=H_0 + \Delta H_s,~~ \Delta H_s = H_{SL} +H_{SS} + H_{T}
\label{1}\ee where $H_0$ is the generalization of the spinless
$q\bar q$ Hamiltonian, obtained in \cite{o19}, and given in
\cite{o17,o18}
$$
H_0=\frac{p_r^2}{\mu(t)}+
\mu(t)+\frac{L(L+1)}{r^2[\mu+2\int^1_0(\beta-\frac12 )^2\nu
d\beta]}+
$$
\be +\int^1_0\frac{\sigma^2_{adj}d\beta}{2\nu(\beta,t)} r^2
+\frac12 \int^1_0 \nu(\beta, t) d\beta.\label{2} \ee Here $\mu(t)$
and $\nu(\beta,t)$ are positive auxiliary functions which are to
be found from the  extremum condition \cite{o17}. Their extremal
values are equal to the  effective gluon energy $\langle
\mu\rangle $ and energy density of the adjoint string $\langle
\nu\rangle $.

Here $\sigma_{adj} =\frac94 \sigma_{fund}$, and we shall always
set $\sigma_f =0.18$ GeV$^2$ as found from meson Regge
trajectories \cite{o17,o29}.

To find the spin-averaged masses from (\ref{1}), $ \Delta H_s =0$,
one can use the WKB procedure developed in \cite{o30} for the
$q\bar q$ case and having accuracy better than 5\% for $n=0$,
which yields the values given in Table 1 (exact values available
for $L=0$ are given in parentheses).

\vspace{1cm}
{\bf Table 1}\\
Spin-averaged masses (in GeV) of $gg$ states with $L=0,...5$ and
$n=0,1,2,~~ \sigma_f =0.18$ GeV$^2$
 \vspace{1cm}

 \begin{tabular}{|ll|l|l|l|l|l|l|} \hline
 &$L$&0&1&2&3&4&5\\
 $n$&&&&&&& \\ \hline
 0&&2.09 (2.01)&2.65&3.13&3.53&3.88&4.21\\ \hline
 1&&3.20 (2.99)& 3.65&4.03&4.37& 4.67&4.95\\     \hline
 2&&4.01 (3.75)&4.40&4.74&5.04&5.31&5.56\\             \hline
 \end{tabular}
\vspace{1.5cm}

In (\ref{1}) and in Table 1 the  effect of perturbative gluon
exchanges between gluons was neglected, and  we consider it as a
leading reasonable approximation in obtaining glueball masses. In
doing so we follow the argument given in \cite{o8}, where it was
shown that the Coulomb-like adjoint charge interaction is not
formed between valence  gluons, and moreover, following BFKL
approach, one can consider gluon exchange as an effectively not
large which can be deduced from the relatively small shift
$\Delta\equiv \alpha_P(0)-1$ of the intercept, when terms
$O(\alpha_s^2)$ are taken into account \cite{12,o8}. Therefore in
the first approximation we neglect perturbative gluon exchanges,
keeping them only in the spin-dependent terms $\Delta H_S$. This
strategy is supported by the comparison with lattice data of the
spin-averaged masses (see Table 4 of ref. \cite{o8}) and,
separately, by the comparison of our spin splittings of masses
with lattice data, to be discussed below.

Spin-splitting terms $\Delta H_s$ are considered in detail in
\cite{o8} for $L=0,1,2$ and we list in Table 2 the resulting
masses,  in comparison with existing lattice data.

The difference  of Table 2 from the corresponding Table 6 of ref.
\cite{o8} is that we fix $\sigma_f =0.18$ GeV$^2$ and recalculate
lattice masses for this value of $\sigma_f$. Moreover, we take for
$L=0$ in brackets $\alpha_s(eff)=0.2$ (for spin-splitting terms,
since the corresponding spin  interaction occurs at relatively
small distances, while for the rest masses $\alpha_s(eff)=0.3$ was
taken, as in \cite{o8}).

\newpage
\vspace{1cm}
{\bf Table 2}\\

Comparison of  calculated glueball masses (in GeV) with lattice
data  ($\sigma_f=0.18$ GeV$^2  $, $\alpha_s=0.3$ ($\alpha_s=0.2$
in parentheses))

\vspace{0.5cm}

 \begin{tabular}{|l|l|l|l|l|} \hline
$J^{PC}$& $M_{theory}$ &
\multicolumn{3}{c|}{$M_{lat}$}\\\cline{3-5}
 &this work& [22]&[23]& [24]\\\hline

$0^{++}$&(1.61) 1.41&1.53$\pm$0.10&1.53$\pm$0.04& 1.52$\pm$0.13\\
$2^{++}$&(2.21) 2.30&2.13$\pm$0.12&2.20$\pm$0.07& 2.12$\pm $0.15\\
$0^{++*}$&(2.72) 2.41 &2.38$\pm$0.25&2.79$\pm$0.09&\\
$2^{++*}$&(3.13) 3.32 &2.93$\pm$0.14&2.85$\pm$0.28&\\
$0^{-+}$&2.28& 2.30$\pm$0.15& 2.11$\pm$0.24&2.27$\pm $0.15   \\
$0^{-+*}$&3.35& 3.24$\pm$0.2&&          \\
$2^{-+}$&2.70&2.76$\pm$0.16&3.0$\pm$0.28& 2.70$\pm $0.19\\
$2^{-+*}$&3.73& 3.46$\pm$0.21&&          \\
\hline
  \end{tabular}
\vspace{0.5cm}

 One can see in Table 2 a good agreement of our
calculated glueball masses with measured lattice values,
especially for ground states. The radially excited states (marked
with  an asterix) are some 5-7\% higher than lattice data, which
probably is the result of WKB approximation \cite{o30} for the
Hamiltonian (\ref{2}); indeed as seen from Table 3 of \cite{o30},
and our Table 1, second column, the exact mass eigenvalues for
$L=0$ and $n_r>0$ are 6-7\% lower than those from WKB
approximation.

This agreement of theory with lattice data  becomes even more
striking, when one  realizes that our calculation has no fitting
parameters at all, since string tension is a given scale parameter
(one could compare dimensionless $M/\sqrt{\sigma}$, as it is done
in Table 4 of \cite{o8}) and $\alpha_s,$ fixed at the
characteristic value, $\alpha_s=0.3$ describes only spin splitting
of masses.

Let us now discuss high spin states, not present in Table 2. The
$L=2, S=2$ states include $0^{++}, 1^{++}, ..., 4^{++}$ states
which are spread over the mass distance of 63 MeV due to the
spin-orbit and tensor level splitting (see table 7 of \cite{o8}).
Here perturbative and nonperturbative (Thomas term) spin-orbit
interaction almost  cancel each other.

One can expect, that for higher states the splitting of the states
will be even less, as it is observed experimentally for mesons and
in what  follows we shall neglect spin splitting for states with
$L>2$.

The resulting glueball masses are given in Table 3.

\vspace{1cm}
{\bf Table 3}\\

Glueball masses (in GeV) for $L\geq 2$, from the Hamiltonian (2)

\vspace{0.5cm}

 \begin{tabular}{|l|l|l|} \hline
$L,S$& States& Masses\\
\hline 2,2&$0^{++},4^{++},1^{++},2^{++},3^{++}$ & 3.11-3.17\\
2,0& $2^{++}$&3.13\\
3,1& $4^{-+}, 3^{-+},2^{-+}$& 3.53\\
4,2 &$6^{++}-2^{++}$&3.88\\
4,0 &$4^{++}$& 3.88\\
5.1&$6^{-+},5^{-+}, 4^{-+}$& 4.21\\ \hline
\end{tabular}

\vspace{1cm}

These masses can be compared to (scarce) lattice values. E.g.
$M_{lat} (3^{++}) = 3.28\pm  0.2$ GeV in \cite{o22}, 3.81$\pm$0.31
in \cite{o23} and 3.26$\pm 0.21$ in \cite{o24}; $M_{lat}(4^{++})
=3.65$ GeV in \cite{o2} and around 3.5 GeV in  \cite{o1}; $M_{lat}
(6^{++})=4.2\pm 0.2)$ GeV in  \cite{o1}.

One can notice an approximate degeneracy of $3^{++}$ and $4^{++}$
states on the lattice in agreement with theory. For the pomeron
trajectory the state $6^{++}(L=4,S=2$) is important and will be
used below.

It follows from Table 3, that there are several states with the
same $J^P$ which do not differ substantially in  mass (for example
$4^{++}$ in (2,2), (4,2)  and (4,0) configurations). This makes
difficult to compare predicted masses for these states with
lattice data, which are usually approximated by a single state.
Note that masses of states with $J^P=4^+ 6^+$ on the pomeron
trajectory calculated in ref \cite{o1} are higher than the lowest
values, obtained in our paper. As a result the slope of the
pomeron trajectory obtained in paper \cite{o1} is smaller than the
 value corresponding to Casimir scaling and an intercept of the
 purely glueball pomeron  is higher than in the present approach.
 As  we discuss in Section 4 below this intercept has little to do
 with a physical pomeron intercept due to large  mixing with $q
 \bar q$ trajectories.

\section{Three-gluon glueballs}

The $3g$ glueballs (oddballs) can be of two basic configurations:
the $\Delta$-type and the $Y$-type (only the first one was
considered in \cite{o8}). The corresponding wave operators for the
$\Delta$-type are given in Table 10 of \cite{o8} and basically
correspond to 3 gluons sitting in vertices  of a triangle and
connected by  the fundamental strings. Another, not  considered in
\cite{o8} the $Y$, form is composed as \be \Psi (x_1,x_2,x_3)\sim
d_{abc} E^a_i(x) E^b_k(x) E^c_l(x)\label{2a}\ee and is the adjoint
equivalent of the baryon operator with the  replacement
$e_{\alpha\beta \gamma} \to d_{abc}$. Another  possible $Y$-type
form is made with the operator $f_{abc}$ instead of $d_{abc}$ and
requires antisymmetric spin-coordinate function. Using the
charge-conjugation  $C$ transformation $F_{\mu\nu}\to
-F_{\mu\nu}^T$, one can easily understand that the form made of
$f_{abc}$ has $C=+1$. Both $f,d$ forms have been used in
\cite{o3}.

The spin-independent part of Hamiltonian in both cases can be
written as \be H_0^{(3g)} = T_{3g} + \frac{3\mu}{2} + V_{Y,\Delta}
(\ver_1, \ver_2, \ver_3)\label{3}\ee where $T_{3g}$ is the kinetic
operator, \be T_{3g} =\frac{\vep^2_\eta+\vep^2_\xi}{2\mu},~~ \vep
=\frac{1}{i} \frac{\partial}{\partial\vexi},~~
\vep_\eta=\frac{1}{i}\frac{\partial}{\partial{\veta}}\label{4}
 \ee
and the Jacobi coordinates defined as
 \be \vexi=\sqrt{\frac{3}{2}}
 \left(\frac{\ver_1+\ver_2}{2}-\ver_3\right),~~
\veta=\frac{\ver_1-\ver_2}{\sqrt{2}}.\label{5}\ee Finally the
interaction is \be V_\Delta =\sigma_f \sum_{i<j}
|\ver_i-\ver_j|,~~ V_Y =\sigma_{adj}\sum^3_{i=1} |\ver_i-
\veR_Y|\label{6}\ee and $\veR_Y$ is the position of the string
junction. We shall be using $\sigma_f\equiv \sigma, ~~\sigma_{adj}
=\frac{C_2(adj)}{C_2(fund)}\sigma = \frac94 \sigma$.

We are solving equation $H_0^{(3g)} \Psi=M(\mu)\Psi$ as in
\cite{o8} using the hyperspherical approach \cite{o31,o32}, which
yields very good accuracy already in the lowest approximation
\cite{o33}. Defining the  hyperradius $\rho, \rho^2=\veta^2
+\vexi^2$, and grand orbital momentum $K,~~ K=L, ~~ L+2, ~~
L+4,..$, one has the equation \be
-\frac{1}{2\mu}\frac{d^2\chi}{d\rho^2} +U_{\Delta, Y}(\rho)
\chi(\rho) =\varepsilon(\mu) \chi(\rho),\label{7} \ee with \be
U_{\Delta, Y}(\rho)
=\frac{1}{2\mu\rho^2}\left(K^2+4K+\frac{15}{4}\right)+ c_{\Delta,
Y} \rho\sigma,\label{8}\ee \be c_{\Delta}=
\frac{32\sqrt{2}}{5\pi}=2.88; ~~ c_Y=3.31.\label{9} \ee Here one
notice that the case of $Y$ form can be obtained from baryonic
calculations of \cite{o32} by a simple replacement $\sigma_f\to
\sigma_{adj}$ (see \cite{o32} for details of derivation) and the
total mass $M(3g)$ is obtained by minimizing the mass $M(\mu)$
over the values of $\mu$, \be M(3g) =\min_\mu M(\mu) =\min_\mu
\left\{ \frac{3\mu}{2} +\varepsilon (\mu)\right\}.\label{10}\ee It
was shown \cite{o33}, that $\varepsilon(\mu)$ can be found (with
one percent accuracy) from the minimum of $U_{\Delta, Y}(\rho)$ at
some point $\rho=\rho_0$.

In this way one obtains for $\varepsilon(\mu)$. \be
\varepsilon_{\Delta, Y}(\mu) =\frac32 \frac{(c_{\Delta,
Y})^{2/3}\sigma^{2/3}}{\mu^{1/3}}  k^{1/3}
\left(1+\frac{1}{\sqrt{3k}}\right) \equiv \frac{\Omega_{\Delta,
Y}\sigma^{2/3}}{\mu^{1/3}}\label{11}\ee where $k\equiv
K^2+4K+\frac{15}{4}$.

Minimizing over $\mu$ in (\ref{10}), one finds the constituent
gluon mass $\mu_0$, \be \mu_0 =\left( \frac29\Omega_{\Delta,
Y}\right )^{3/4}\sqrt{\sigma},~~ M(3g) =6\mu_0.\label{12}\ee

One can  now predict the $3g$ glueball masses, still without spin
splittings, fixing the value of $K_{min}=L=0,1,2,...$

For the lowest $3g$ state  with $K_{\min}=L=0,$ the spin splitting
is due to the hyperfine interaction and it  was calculated in
\cite{o8}, yielding \be \Delta M_{ss} =0.644 \mu_0
\frac{J(J+1)-6}{6}\label{13} \ee

This gives $\Delta M_{ss}(3^{--})=0.28$ GeV, $\Delta M_{ss}
(2^{--})=0,~~ \Delta M_{ss} (1^{--})=-0.189$ GeV.

For $L>0$ splitting is due to tensor and spin-orbit forces, and we
assume, that the situation is similar to that of baryons, where
these forces are known to be weak. The situation might however be
different for $3g$ states.

We neglect spin splittings for $L>0$ and give in Table 4 the
calculated  spin-averaged masses. One should note that the
$Y$-type glueballs are fully equivalent to baryons and their
masses are obtained by simply multiplying baryon masses from
\cite{o32}  by the factor
$\sqrt{\frac{C_2(adj)}{C_2(fund)}}=\frac32$ More sophisticated
configurations were also considered in \cite{o32} however the
resulting masses are very close to the corresponding $M_Y$ values
and are omitted from the Table 4.

\vspace{1cm}
{\bf Table 4}\\

Oddball masses (in GeV) for various $L=0,1,...4$ and $J\leq 7$ in
comparison with lattice results. For $L>0$ only spin-averaged
values are shown, $\sigma=0.18$ GeV$^2$.

\vspace{0.5cm}

 \begin{tabular}{|l|l|l|l|ll|} \hline
$L$& $J^{PC}$&$ M_{\Delta}(theor) $& $M_Y(theor)$ & $M(lat)$&\\
&&this work &this work&[24]&[1]\\ \hline 0&$1^{--}$&3.02&3.32&3.40$\pm 0.21$ &3.10\\
 &$2^{--}$&3.21&3.53&3.56$\pm 0.21$ &3.55\\
 &$3^{--}$&3.49&3.83&3.73$\pm 0.21$ &4.15\\
 1&$0^{+-},...3^{+-}$&3.72&4.09&3.46$\pm 0.2$ &$\sim$3.2\\
2&$5^{--},...1^{--}$&4.18&4.59&&\\
4&$7^{--},...1^{--}$&4.96&5.25&&\\
\hline
\end{tabular}

\vspace{1cm}

One can see from Table 4, that agreement between theoretical and
lattice values is reasonable and of the same quality, as the
agreement between results of different lattice groups. Our results
are closer to the lattice data of \cite{o22,o24} and lie below
those of \cite{o1}; the spin splittings in the ($1^{--}, 2^{--},
3^{--}$) triplet are 0.47 GeV in our calculation \cite{o8} and
0.33$\pm$ 0.21 in data of \cite{o22,o24}, which again may indicate
that effective $\alpha_s$ for spin-spin interaction is
$\alpha_s=0.2$ rather than our fixed value $\alpha_s=0.3$.

Two other models have been used in \cite{o3} for $1^{--}, 2^{--},
3^{--}, 5^{--}, 7^{--}$ states with results similar to ours for
the first triplet of states, but much heaver for $5^{--}, 7^{--}$
states.

\section{Pomeron and odderon trajectories}

Since our results are the same as in \cite{o8} for $L=0,2$, and
the resulting pomeron trajectory is assumed to be the same as in
\cite{o8} (mixed with $f,f'$ trajectories), it is worthwhile to
compare the slopes $\alpha_p'(0)$ obtained from the standard
procedure, $\alpha_p' (stand) =\frac{1}{2\pi\sigma_{adj}}=0.393$
GeV$^{-2}$, with the slopes obtained from masses $M(J=2^{++};
4^{++}; 6^{++})=(2.3-2.21; 3.15; 3.88)$ GeV, where we give the
$2^{++}$ mass interval for $\alpha_s=0.3-0.2$.

One obtains for the neighboring masses \be
\alpha_P'(4^{++}-2^{++})= (0.431\div 0.397){\rm ~
GeV}^2,~~\alpha_P'(6^{++}-4^{++})= 0.3971{\rm ~
GeV}^2.\label{14}\ee

Thus we see that the slopes are close to the standard one, and the
trajectory is close to the straight line. The value of intercept,
on the other hand depends crucially on the intersection with $f,
f'$ trajectories and as we argued in \cite{o8}, this value is not
actually controlled by the masses on  the pomeron trajectory. This
is  in contrast to the case of odderon, where no intersection with
meson trajectories is possible for $t>0$ and hence intercept can
be estimated from the computed above masses, and now we have three
states on the trajectory and can determine both slope and
intercept.

We  define two odderon trajectories corresponding to the
$\Delta$-type and $Y$-type  configurations,

\be \alpha_\Delta(t)=\alpha'_\Delta\cdot t + \alpha_\Delta (0),~~
\alpha_Y(t)=\alpha'_Y\cdot t + \alpha_Y (0).\label{16}\ee

It appears that one can find odderon trajectory analytically,
using explicit mass formulas Eqs. (\ref{11})-(\ref{13}). Here spin
splittings are neglected, as it usually done for meson Regge
trajectories \cite{o29}.
 E.G. in the latter case spin dependent correlations split the
 spin-averaged Regge trajectory into the two  nonlinear trajectories with
 $J=L\pm 1$.

 In this way one can approximate $k$ in (\ref{12}) for $K=L$, as
 $k=(L+2)^2-\frac{1}{4}\approx (L+2)^2$ and hence
 \be
 M^2_\Delta(3g)/\sigma = 4c_\Delta \sqrt{3k} \left(
 1+\frac{1}{\sqrt{3k}}\right)^{3/2} \approx 20.0
 \left(L+2+\frac{\sqrt{3}}{2}\right).\label{17}\ee

 Taking into account, that $J_{odd} =L+3$, one obtains the odderon
 trajectory
 \be M^2(3g) =20.0\sigma (J-\alpha_\Delta (0)) ,~~
 \alpha_\Delta(0) =1-\frac{\sqrt{3}}{2}=0.134.\label{18}\ee
 At this point one can compare oddball masses from (\ref{18}) for
 $J=3^{--}, 5^{--}$ and $7^{--}$ with the values given in the
 Table 4, and find that these masses  differ less than 1\%, if for
 $M(3^{--})$ one takes the spin-averaged value, almost coinciding
 with $M(2^{--})=3.21$.

 It is interesting to compare the resulting slope in (\ref{18})
 $\alpha^\prime_\Delta =\frac{1}{20\sigma} =0.277$ GeV$^{-2}$ with
 the standard relativistic "potential" slope (when string rotation
 correction is not taken into account which is our
 case),$\alpha^\prime_\Delta(pot) =\frac{1}{8\sigma_a} =0.308$
 GeV$^{-2}$. The string correction was computed in \cite{o17, o29}
  and it decreases the glueball masses with $L\neq 0$ by $4\div
  5\%$ and the asymptotic $(L\gg 1)$ slope is
  $\alpha^\prime_\Delta(string)=\frac{1}{2\pi\sigma_a}=0.39$
  GeV$^{-2}$. If one uses this slope (and keeps the mass with
  $L=0$ unchanged), then one has the string-corrected odderon
  trajectory
  \be M^2_{string} (3g) = 2 \pi \sigma_a (J-\alpha_\Delta^{string}
  (0) ), ~~ \alpha_\Delta^{string} (0) =- 1.05.\label{19}\ee

 It is clear from (\ref{10}), (\ref{12}), that masses of $Y$-type
 glueballs are higher, namely
 $ \frac{M_Y}{M_\Delta} =\sqrt{\frac{c_Y}{c_\Delta}}\cong 1.1$,
 and  the intercept of $Y$ trajectory  is lower.

 The same  procedure as for Eq. (\ref{19}) leads for the intercept of $Y$-type trajectory the value
 \be \alpha_Y^{string}(0)=-1.9.\label{20}\ee
Now if one insists on using spin-splitted (nonlinear in general)
trajectory passing through $3^{--}$ and $5^{--}$ states, one
obtains extrapolating $\alpha_\Delta(t)$ to $t=0$ the intercept
$\alpha_\Delta =-1.6.$

We see that determination of intercepts of  $3g$-trajectories is
model dependent, but  in
  both cases, Eq. (\ref{18}) and Eq. (\ref{19}), the odderon
  intercept is far from unity.

\section{Discussion and summary}

As shown in the previous section, the  odderon intercept is 0.14
(without string corrections) and drops down to -1.05 when  the
string slope  $\frac{1}{2\pi\sigma_a}$ is restored.
 In any case
intercepts of these $3g$-trajectories   are very far from the
perturbative intercept $\alpha_{3g}\approx 1$. Thus the
nonperturbative effects strongly reduce the intercept of the
$3g$-trajectories and there is no "odderon" (the singularity with
negative  $C=\sigma $ and $\alpha_o(0)\approx 1$) in our approach.

This difference for the intercepts of $3g$-singularities with
results of perturbation theory is of principle importance and
takes place for all multigluonic states. In the perturbation
theory the multigluon singularities in $j$-plane are above unity
\cite{a} and in principle mix with $2g$-state. Account of
nonperturbative interaction between gluons lead to large masses of
multigluon states and  consequently to low intercepts of such
trajectories. Qualitatively it can be understood as follows:
 NP interactions confine and create effective mass for each gluon (calculable in FCM \cite{o15, o16})  proportional to $\sqrt{\sigma}$.
 Therefore $4g$ states are separated from $2g$ states by an
 interval $\sim 2$ GeV and mixing can be  neglected in the  first
 approximation. This is contrast to BFKL method, where such
 interval is absent and in principle all multigluon states should
 be considered simultaneously.

 Comparison of our calculated masses with lattice data in Tables
 2,3 shows that the ordering  of states in mass values is the
 same. The agreement in masses for states with $L=0,1$ and $n_r=0$
 is surprisingly good although no fitting parameters are used.
 The (small) discrepancy for the states with $n_r>0$ was discussed above,
 and it is planned to improve theoretical accuracy for $L>0,
 n_r>0$. For $L\geq 2$ our spin-averaged
 values  are in a good agreement with lattice data of \cite{o22,o24} and   10$\div
 $20\% below data of \cite{o1, o2, o23}.
 One should take into account at this point, that $4^{++}$ and
 $6^{++}$ states occur from $L=2$ and 4, and $L=4$ and 6
 respectively, and resulting mixing can shift the masses from the
 one-channel values in lattice computations (as well as in our analytic approach).

 For $3g$ spin-averaged states in Table 3 the agreement with
 lattice data from \cite{o24, o1} is within 10\%.

 Our results for the odderon masses of $J^{PC}= 5^{--}, 7^{--}$
 are some 0.5 GeV below the results of models $H^g_{eff}$ and
 $H_M$ from \cite{o3}, however qualitatively  the odderon
 intercepts are also low (-0.88 and 0.25  respectively).

 In summary, we have calculated high spin states of both $2g$ and
 $3g$ systems and found the corresponding masses, in this way
 extending our results from an earlier paper \cite{o8}. We have
 confirmed our previous results on the pomeron
 trajectoriy.

 In the   case of odderon  we  have found two trajectories, of $\Delta$
 and $Y$ type  with masses differing by $\sim 10\%$ and similar
 intercepts  below 0.14. This result might explain why odderon is
 not  seen in photonucleon  reactions \cite{o6, o7}, and  is in
 sharp contrast to the BFKL-type odderon
 intercept, which is around 1.   This  work  is supported
  by the Federal Program of the Russian Ministry of industry, Science and Technology
  No.40.052.1.1.1112, and by the
grant for scientific schools NS-1774. 2003. 2. One of authors (A.B.K.) acknowledges  partial support of the grants CRDF RUP2-2621-M0-04
and 04-02-17263. \\

\end{document}